\documentclass[structabstract]{aa}
\usepackage{txfonts}
\usepackage{graphicx}
\usepackage{aalongtable}
\begin{document}
\title{Rotational spectroscopy of AlO}
\subtitle{Low-\textit{N} transitions of astronomical interest in the \textit{X}\textsuperscript{2}$\Sigma$\textsuperscript{+} state}
\author{O. Launila\textsuperscript{1}
\and D.P.K. Banerjee\textsuperscript{2}}
\institute{KTH AlbaNova University Center,
Department of Applied Physics,
SE-106 91 Stockholm, SWEDEN,
E-mail: olli@kth.se
\and Astronomy and Astrophysics Division,
Physical Research Laboratory,
Ahmedabad 380009,  Gujarat, INDIA,
E-mail: orion@prl.res.in}
\date{Received 9 September 2009/ Accepted 9 October 2009}
\abstract
{}{The detection of rotational transitions of the AlO radical at millimeter wavelengths from an astronomical source has recently been reported. In view of this, rotational transitions in the ground \textit{X}\textsuperscript{2}$\Sigma$\textsuperscript{+} state of AlO have been reinvestigated. }
{Comparisons between Fourier transform and microwave data indicate a discrepancy regarding the derived value of $\gamma$$_{D}$ in the v = 0 level of the ground state. This discrepancy is discussed in the light of comparisons between experimental data and synthesized rotational spectra in the v = 0, 1 and 2 levels of \textit{X}\textsuperscript{2}$\Sigma$\textsuperscript{+}.}
{A list of calculated rotational lines in v = 0, 1 and 2 of the ground state up to \textit{N}' = 11 is presented which should aid astronomers in analysis and interpretation of observed AlO data and also facilitate future searches for this radical.}{}
\keywords{Molecular data--Molecular processes--Interstellar medium: molecules--Radiolines: stars}
\maketitle
\titlerunning{Rotational transitions of AlO}
\authorrunning{O. Launila and D.P.K Banerjee}
\section{INTRODUCTION}
\paragraph{ The ground \textit{X}\textsuperscript{2}$\Sigma$\textsuperscript{+} state of the AlO radical has been studied with microwave spectroscopy in the v = 0, 1 and 2 levels. T\"{o}rring et al. (1989) recorded the \textit{N} = 1
$\rightarrow$2 transition near 76 GHz. In their Table 1, frequencies for several $\Delta$\textit{F} = $\Delta$\textit{N} and $\Delta$\textit{F} $\neq$ $\Delta$\textit{N} transitions are shown. Yamada et al. (1990) and Goto et al. (1994) recorded several $\Delta$\textit{F} = $\Delta$\textit{N} rotational transitions, resulting in a set of accurate molecular constants, including $\gamma$ and $\gamma$$_{D}$. }
\paragraph{  Launila et al. (1994) have performed a Fourier transform study of the \textit{A}\textsuperscript{2}$\Pi$$_{i}$\textbf{$\rightarrow$}\textit{X}\textsuperscript{2}$\Sigma$\textsuperscript{+} transition of AlO in the 2 $\mu$m region. In that work, some discrepancies were pointed out regarding their derived $\gamma$$_{D}$ values, as compared to those found in the microwave work. While Yamada et al. (1990) had given a positive value for $\gamma$$_{D}$ for v = 0, the sign was in fact found to be negative in the light of high-\textit{N} data of Launila et al. (1994). One of the aims of the present paper is to reinvestigate this discrepancy more closely. The work by Goto et al. (1994), dealing with the v = 1,2 vibrational levels of the ground state of AlO, does not show the same discrepancy, however. }
\paragraph{In a theoretical work, Ito et al. (1994) have discussed and explained the observed vibrational anomalies in the spin-rotation constants of the ground state in the light of spin-orbit interaction with the \textit{A}\textsuperscript{2}$\Pi$$_{i}$ and \textit{C}\textsuperscript{2}$\Pi$ states.}
\paragraph{Recently, Tenenbaum and Ziurys (2009) reported three rotational transitions of AlO from the supergiant star VY Canis Majoris. In order to facilitate future millimeter-wave search for rotational transitions of AlO, tables containing expected frequencies in the v = 0, 1 and 2 levels of the ground state are useful. In the present work, such tables are presented.}
\section{ANALYSIS AND RESULTS}
\paragraph{The \textit{X}\textsuperscript{2}$\Sigma$\textsuperscript{+} ground state of the AlO radical represents a good example of a Hund's case (b$\beta$$_{S}$) coupling. Here, the nuclear and molecular spins \textbf{I} and \textbf{S} couple to form an intermediate vector \textbf{G}, which subsequently couples with \textbf{N} to form \textbf{F}. Although the energy levels can be calculated with a standard hyperfine Hamiltonian based on \textbf{J} = \textbf{N}+\textbf{S} formalism, the \textit{J} quantum number is to be considered as a bookkeeping device only. This is the case in Table I of Yamada et al. (1990), where both \textit{J}' and \textit{J}" have been included. Goto et al. (1994) only specify the 'good' quantum numbers \textit{N} and \textit{F} in their Tables 1 and 2.}
\paragraph{In the work of Launila et al. (1994) it was found that the positive sign of $\gamma$$_{D}$ for v = 0 was in conflict with the high-\textit{N} behavior of the spin doubling of the ground state, as shown in Figs 3 and 4 of their study. We have plotted in the present work the observed \textsuperscript{R}\textit{Q}$_{21}$ and \textit{R}$_{2}$ branches in the (2,0) band of the \textit{A}\textsuperscript{2}$\Pi$$_{i}$$\rightarrow$\textit{X}\textsuperscript{2}$\Sigma$\textsuperscript{+} transition from \textit{N} = 3 to \textit{N} = 104 (Fig. 1). Here, we are using standard spectroscopic notations, where superscripts in branch designations refer to \textit{N}-numbered branch types, while subscripts refer to the spin components. For instance, \textsuperscript{R}\textit{Q}$_{21}$ denotes a \textit{Q}-branch of \textit{R}-type, going from the upper state spin component 2 to the lower state component 1. Vibrational bands are denoted as (v$_{upper}$,v$_{lower}$). From the plot in Fig. 1 we can see that the spin doubling of the ground state v = 0 level increases with \textit{N}, up to about 0.1 cm\textsuperscript{-1} for \textit{N} = 50, after which it starts to decrease until it finally reaches the value of about 0.07 cm\textsuperscript{-1} for \textit{N} = 104. According to the constants of Yamada et al. (1990), the spin doubling would have to increase monotonically to about 0.3 cm\textsuperscript{-1}, which is in contrast with the observations.}

\begin{figure}[h]
\centering
\includegraphics[scale=0.32]{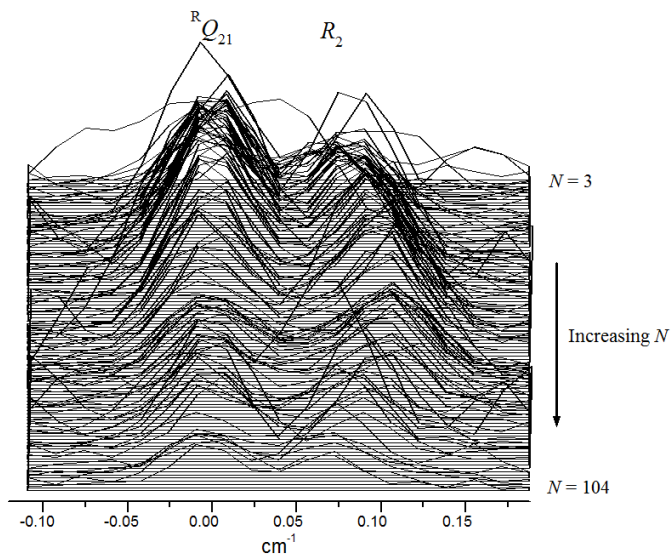}
\caption {The \textsuperscript{R}\textit{Q}$_{21}$ and \textit{R}$_{2}$ branches in the (2,0) band of the \textit{A}\textsuperscript{2}$\Pi$$_{i}$$\rightarrow$\textit{X}\textsuperscript{2}$\Sigma$\textsuperscript{+} transition of AlO. The traces show a series of 0.3 cm\textsuperscript{-1} intervals, cut directly from the FT spectrum. The \textsuperscript{R}Q$_{21}$ lines have been arranged so that their reduced wavenumbers coincide. Some lines belonging to other branches are also visible in the plot.}
\end {figure}

\paragraph{We have also recalculated the rotational transition \textit{N} = 10$\leftarrow$9 shown in Fig. 2 of Yamada et al. (1990), using the constants given by them ($\gamma$ = 51.660 MHz, $\gamma$$_{D}$ = 0.00343 MHz), and also using the same constants, but with a reversed sign of $\gamma$$_{D}$. The Hamiltonian used was the same as in Launila et al. (1994). The results are presented in Fig. 2.}

\begin{figure}[h]
\centering
\includegraphics[scale=0.28]{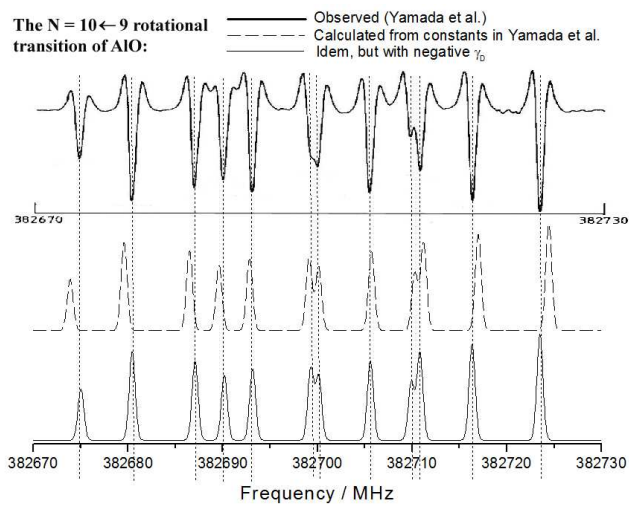}
\caption {The \textit{N} = 10\textbf{$\leftarrow$}9 rotational transition in the \textit{X}\textsuperscript{2}$\Sigma$\textsuperscript{+} (v=0) ground state of AlO from Yamada et al. (1990), as compared with calculated spectra convoluted with a Gaussian line profile of 0.75 MHz FWHM.}
\end {figure}

\paragraph{The uppermost trace of Fig. 2 shows the experimental data, while the middle one shows the recalculated rotational spectrum using the constants of Yamada et al. (1990). The lower trace shows the same spectrum calculated with reversed sign of $\gamma$$_{D}$. The calculated spectra have been convoluted with a Gaussian profile of 0.75 MHz FWHM. It is clear from this comparison that the reversed sign of $\gamma$$_{D}$ results in an almost perfect agreement with the experimental data, while the unaltered constants by Yamada et al. (1990) give rise to deviations of up to $\pm$1 MHz. Our conclusion is that the data fit in the work of Yamada et al. (1990) is essentially correct, although the sign of $\gamma$$_{D}$ has to be changed. This sign change alone results in overall deviations of less than $\pm$0.1 MHz. However, no new least-squares fit of the data in Yamada et al. (1990) has been performed in the present work. The intensities in the calculated spectra were derived using an intensity formula (B1) from Bacis et al. (1973). Although this formula had actually been written for a Hund's case (b$\beta$$_{J}$) - (b$\beta$$_{J}$) transition, the agreement with experimental data is surprisingly good. }

\paragraph{The calculated frequencies and relative intensities (at 230 K) for all $\Delta$\textit{N} = $\Delta$\textit{F} rotational transitions in v = 0,1 and 2 up to \textit{N}' = 11 are given in Tables 1-3.}

\section{DISCUSSION}

\paragraph{The determination of the correct constants for a molecule/radical has its own intrinsic value. Additionally, the present study is also relevant in an astronomical context. Tenenbaum and Ziurys (2009) have recently made the first radio/mm detection of the  AlO (\textit{X}\textsuperscript{2}$\Sigma$\textsuperscript{+}) radical toward the envelope of the oxygen rich supergiant star VY Canis Majoris (VY CMa). They observed the \textit{N} = 7$\rightarrow$6 and 6$\rightarrow$5 rotational transitions of AlO at 268 and 230 GHz and the N = 4\textbf{$\rightarrow$}3 line at 153 GHz. While their search for the \textit{N} = 7$\rightarrow$6 hyperfine transitions was based on direct laboratory measured frequencies by Yamada et al  (1990), the search for the \textit{N} = 6$\rightarrow$5 and \textit{N} = 4$\rightarrow$3 transitions was based on frequencies calculated from spectroscopic constants of those authors. As has been pointed out, an error is present in one of these constants ($\gamma$$_{D}$ in the v = 0 level of the ground state), which leads to deviations between the calculated and true frequencies. While these deviations are small and may not affect the final result of a line search too seriously, it is still desirable to have accurate frequencies to  facilitate future millimeter-wave searches for rotational transitions of AlO. It is planned for several such searches to be taken place in the near future as a consequence of the recent detection in VY CMa. AlO is slowly emerging as a molecule which could attract a fair deal of interest among astronomers. In the VY CMa detection for example it is shown how its study could lead to a better understanding of the gas-phase refractory chemistry in oxygen-rich envelopes. AlO has also been proposed to be a potential molecule in the formation of alumina, which is one of the earliest and most vital dust condensates in oxygen rich circumstellar environments (Banerjee et al. 2007). The mineralogical dust condensation sequence and processes involved are issues of considerable interest to astronomers. The AlO radical also received considerable attention after the strong detection of several \textit{A}$\rightarrow$\textit{X} bands in the near-infrared (1 - 2.5 microns) in the eruptive variables V4332 Sgr and V838 Mon (Banerjee et al. 2003; Evans et al 2003). Other IR detections of AlO include IRAS 08182-6000 and IRAS 18530+0817 (Walker et al. 1997). In the optical, the \textit{B}$\rightarrow$\textit{X} bands have been prominently detected in U Equulei (Barnbaum et al. 1996) and in several cool stars and Mira variables including Mira itself (Keenan et al. 1969;  Garrison 1997).}
\paragraph{The line lists presented here should also aid in analysis of mm line data for aspects related to kinematics. Since the rotational levels of AlO species are split by both fine and hyperfine interactions, each rotational transition consists of several closely spaced hyperfine components. Figure 2 exemplifies this. If line broadening is small and the spectral resolution of the observations is adequate to resolve these components, then different components will yield different Doppler velocities if the corresponding reference or rest frequencies are in error.  This could lead to ambiguity in interpreting the data. Even if the hyperfine components are not distinctly resolved but rather blended to give a composite line profile (as in the observed profiles in VY CMa), modelling of such composite profiles using wrong rest frequencies could lead to errors in estimating  the composite line centre, half width of the composite profile and half-widths of the individual hyperfine components. Proper estimates of such kinematic parameters are important as they help determine the size and site of origin of a molecular species. The detailed discussion by Tenenbaum and Ziurys (2009) in the case of VY CMa, which has three distinctly different kinematic flows in the system, illustrates this.}

\section{SUMMARY}
\paragraph{Rotational transitions of AlO have been reinvestigated. The discrepancy regarding the derived value of $\gamma$$_{D}$ of the \textit{X}\textsuperscript{2}$\Sigma$\textsuperscript{+} ground state v = 0 level has been shown to be due to a sign error in the microwave work by Yamada et al. (1990). A list of calculated rotational lines in v = 0, 1 and 2 of the ground state up to \textit{N}' = 11 has been presented.}

\section{REFERENCES}
\paragraph{Bacis, R., Collomb, R. and Bessis, N. 1973, Phys.Rev. A, 8(5), 2255 \\
Banerjee, D. P. K., Misselt, K. A., Su, K. Y. L., Ashok, N. M., Smith, P. S. 2007, ApJ, 666, L25 \\
Banerjee, D. P. K., Varricatt, W. P., Ashok, N. M. and Launila, O. 2003, ApJ, 598, L31 \\
Barnbaum, C., Omont, A. and Morris, M. 1996, A\&A, 310, 259 \\
Evans, A., Geballe, T. R., Rushton, M. T., Smalley, B., van Loon, J. Th., Eyres, S. P. S., Tyne, V. H. 2003, MNRAS, 343, 1054 \\
Garrison, R. F. 1997, JAAVSO, 25, 70 \\
Goto, M., Takano, S., Yamamoto, S., Ito, H. and Saito, S. 1994, Chem.Phys.Lett. 227, 287 \\
Ito, H. and Goto, M. 1994, Chem.Phys.Lett. 227, 293 \\
Keenan, P. C., Deutsch, A. J. and Garrison, R. F. 1969, ApJ, 158, 261 \\
Launila, O. and Jonsson, J. 1994, J.Mol.Spectrosc. 168, 1 \\
Tenenbaum, E. D. and Ziurys, L. M. 2009, ApJ 694, L59 \\
T\"{o}rring, T. and Herrmann, R. 1989, Mol.Phys. 68(6), 1379 \\
Walker, H. J., Tsikoudi, V., Clayton, C. A., Geballe, T., Wooden, D. H., Butner, H. M. 1997, A\&A, 323, 442 \\
Yamada, C., Cohen, E.A., Fujitake, M. and Hirota, E. 1990, J.Chem.Phys. 92(4), 2146 \\
}

\begin{longtable}{cccccrr}
\caption{Calculated frequencies and intensities of $\Delta$\textit{F} = $\Delta$\textit{N} rotational transitions in the \textit{X}\textsuperscript{2}$\Sigma$\textsuperscript{+} (v=0) ground state of AlO. The constants from Yamada et al. (1990) have been used, with the exception that the sign of $\gamma$$_{D}$ has been reversed. The intensities are calculated according to a Hund's case (b$\beta$$_{J}$) - (b$\beta$$_{J}$) formula (B1) from Bacis et al. (1973), at a temperature of 230 K.}\\
\hline\hline
G & N' & F' & N" & F" & Freq/MHz & Int. (230 K) \\
\hline
\endfirsthead
\caption{continued.}\\
\hline\hline
G & N' & F' & N" & F" & Freq/MHz & Int. (230 K) \\
\hline
\endhead
\hline
\endfoot
    2 &  1 &  3 &  0 &  2 &  38 278.080 &    0.0   \\       
    3 &  1 &  4 &  0 &  3 &  38 281.977 &    6.0   \\       
    2 &  2 &  2 &  1 &  1 &  76 502.453 &    0.0   \\       
    2 &  2 &  4 &  1 &  3 &  76 553.373 &   11.8   \\       
    3 &  2 &  4 &  1 &  3 &  76 559.315 &   17.7   \\       
    2 &  2 &  3 &  1 &  2 &  76 568.067 &    4.1   \\       
    3 &  2 &  5 &  1 &  4 &  76 579.518 &   28.9   \\       
    3 &  2 &  3 &  1 &  2 &  76 677.262 &    9.4   \\       
    2 &  3 &  5 &  2 &  4 & 114 831.410 &   43.0   \\       
    2 &  3 &  2 &  2 &  1 & 114 832.298 &    5.5   \\       
    2 &  3 &  3 &  2 &  2 & 114 835.198 &   14.1   \\       
    2 &  3 &  4 &  2 &  3 & 114 841.162 &   26.4   \\       
    2 &  3 &  1 &  2 &  0 & 114 846.559 &    0.0   \\       
    3 &  3 &  5 &  2 &  4 & 114 850.440 &   55.2   \\       
    3 &  3 &  6 &  2 &  5 & 114 865.239 &   76.2   \\       
    3 &  3 &  4 &  2 &  3 & 114 866.448 &   38.4   \\       
    3 &  3 &  3 &  2 &  2 & 114 888.139 &   25.1   \\       
    3 &  3 &  2 &  2 &  1 & 114 899.206 &   15.1   \\       
    2 &  4 &  6 &  3 &  5 & 153 108.310 &  100.8   \\       
    3 &  4 &  1 &  3 &  0 & 153 109.961 &   10.3   \\       
    2 &  4 &  5 &  3 &  4 & 153 117.008 &   73.1   \\       
    2 &  4 &  4 &  3 &  3 & 153 118.588 &   50.7   \\       
    2 &  4 &  3 &  3 &  2 & 153 121.600 &   33.2   \\       
    2 &  4 &  2 &  3 &  1 & 153 132.878 &   19.9   \\       
    3 &  4 &  2 &  3 &  1 & 153 133.226 &   38.8   \\       
    3 &  4 &  6 &  3 &  5 & 153 133.628 &  121.9   \\       
    3 &  4 &  5 &  3 &  4 & 153 135.760 &   94.0   \\       
    3 &  4 &  4 &  3 &  3 & 153 141.151 &   71.0   \\       
    3 &  4 &  3 &  3 &  2 & 153 142.358 &   52.6   \\       
    3 &  4 &  7 &  3 &  6 & 153 145.778 &  155.0   \\       
    3 &  5 &  2 &  4 &  1 & 191 379.947 &   48.1   \\       
    2 &  5 &  7 &  4 &  6 & 191 382.536 &  192.2   \\       
    2 &  5 &  6 &  4 &  5 & 191 390.676 &  151.2   \\       
    2 &  5 &  5 &  4 &  4 & 191 394.726 &  116.6   \\       
    3 &  5 &  3 &  4 &  2 & 191 398.164 &   96.1   \\       
    2 &  5 &  4 &  4 &  3 & 191 399.767 &   88.1   \\       
    3 &  5 &  4 &  4 &  3 & 191 406.813 &  119.3   \\       
    3 &  5 &  5 &  4 &  4 & 191 409.199 &  148.4   \\       
    3 &  5 &  6 &  4 &  5 & 191 409.272 &  183.5   \\       
    2 &  5 &  3 &  4 &  2 & 191 410.177 &   65.3   \\       
    3 &  5 &  7 &  4 &  6 & 191 411.477 &  224.7   \\       
    3 &  5 &  8 &  4 &  7 & 191 422.055 &  272.3   \\       
    3 &  6 &  3 &  5 &  2 & 229 648.771 &  114.4   \\       
    2 &  6 &  8 &  5 &  7 & 229 653.036 &  324.2   \\       
    2 &  6 &  7 &  5 &  6 & 229 660.782 &  267.5   \\       
    3 &  6 &  4 &  5 &  3 & 229 664.971 &  186.9   \\       
    2 &  6 &  6 &  5 &  5 & 229 665.912 &  218.4   \\       
    2 &  6 &  5 &  5 &  4 & 229 671.899 &  176.7   \\       
    3 &  6 &  5 &  5 &  4 & 229 673.641 &  221.8   \\       
    3 &  6 &  6 &  5 &  5 & 229 677.782 &  264.1   \\       
    3 &  6 &  7 &  5 &  6 & 229 680.325 &  313.5   \\       
    2 &  6 &  4 &  5 &  3 & 229 681.974 &  142.0   \\       
    3 &  6 &  8 &  5 &  7 & 229 684.329 &  370.3   \\       
    3 &  6 &  9 &  5 &  8 & 229 693.846 &  434.8   \\       
    3 &  7 &  4 &  6 &  3 & 267 913.912 &  216.3   \\       
    2 &  7 &  9 &  6 &  8 & 267 918.911 &  503.3   \\       
    2 &  7 &  8 &  6 &  7 & 267 926.336 &  428.6   \\       
    3 &  7 &  5 &  6 &  4 & 267 929.255 &  317.9   \\       
    2 &  7 &  7 &  6 &  6 & 267 931.998 &  362.9   \\       
    3 &  7 &  6 &  6 &  5 & 267 938.051 &  366.9   \\       
    2 &  7 &  6 &  6 &  5 & 267 938.489 &  305.6   \\       
    3 &  7 &  7 &  6 &  6 & 267 943.172 &  424.6   \\       
    3 &  7 &  8 &  6 &  7 & 267 947.009 &  490.7   \\       
    2 &  7 &  5 &  6 &  4 & 267 948.413 &  256.7   \\       
    3 &  7 &  9 &  6 &  8 & 267 951.860 &  565.4   \\       
    3 &  7 & 10 &  6 &  9 & 267 960.593 &  649.0   \\       
    3 &  8 &  5 &  7 &  4 & 306 174.001 &  360.4   \\       
    2 &  8 & 10 &  7 &  9 & 306 179.316 &  735.8   \\       
    2 &  8 &  9 &  7 &  8 & 306 186.459 &  641.1   \\       
    3 &  8 &  6 &  7 &  5 & 306 189.028 &  495.5   \\       
    2 &  8 &  8 &  7 &  7 & 306 192.386 &  556.4   \\       
    3 &  8 &  7 &  7 &  6 & 306 197.956 &  561.0   \\       
    2 &  8 &  7 &  7 &  6 & 306 199.154 &  481.4   \\       
    3 &  8 &  8 &  7 &  7 & 306 203.668 &  636.4   \\       
    3 &  8 &  9 &  7 &  8 & 306 208.233 &  721.4   \\       
    2 &  8 &  6 &  7 &  5 & 306 209.009 &  416.0   \\       
    3 &  8 & 10 &  7 &  9 & 306 213.497 &  816.2   \\       
    3 &  8 & 11 &  7 & 10 & 306 221.612 &  921.0   \\       
    3 &  9 &  6 &  8 &  5 & 344 428.006 &  553.1   \\       
    2 &  9 & 11 &  8 & 10 & 344 433.426 & 1028.0   \\       
    2 &  9 & 10 &  8 &  9 & 344 440.311 &  911.0   \\       
    3 &  9 &  7 &  8 &  6 & 344 443.008 &  726.3   \\       
    2 &  9 &  9 &  8 &  8 & 344 446.356 &  805.1   \\       
    3 &  9 &  8 &  8 &  7 & 344 452.051 &  810.4   \\       
    2 &  9 &  8 &  8 &  7 & 344 453.276 &  710.3   \\       
    3 &  9 &  9 &  8 &  8 & 344 458.123 &  905.6   \\       
    2 &  9 &  7 &  8 &  6 & 344 463.087 &  626.2   \\       
    3 &  9 & 10 &  8 &  9 & 344 463.112 & 1011.7   \\       
    3 &  9 & 11 &  8 & 10 & 344 468.561 & 1128.8   \\       
    3 &  9 & 12 &  8 & 11 & 344 476.168 & 1257.0   \\       
    3 & 10 &  7 &  9 &  6 & 382 675.021 &  800.6   \\       
    2 & 10 & 12 &  9 & 11 & 382 680.433 & 1385.7   \\       
    2 & 10 & 11 &  9 & 10 & 382 687.074 & 1244.3   \\       
    3 & 10 &  8 &  9 &  7 & 382 690.169 & 1016.2   \\       
    2 & 10 & 10 &  9 &  9 & 382 693.151 & 1115.3   \\       
    3 & 10 &  9 &  9 &  8 & 382 699.297 & 1121.2   \\       
    2 & 10 &  9 &  9 &  8 & 382 700.137 &  998.3   \\       
    3 & 10 & 10 &  9 &  9 & 382 705.589 & 1238.4   \\       
    2 & 10 &  8 &  9 &  7 & 382 709.915 &  893.4   \\       
    3 & 10 & 11 &  9 & 10 & 382 710.818 & 1367.7   \\       
    3 & 10 & 12 &  9 & 11 & 382 716.327 & 1509.1   \\       
    3 & 10 & 13 &  9 & 12 & 382 723.495 & 1662.8   \\       
    3 & 11 &  8 & 10 &  7 & 420 914.194 & 1109.0   \\       
    2 & 11 & 13 & 10 & 12 & 420 919.531 & 1814.6   \\       
    2 & 11 & 12 & 10 & 11 & 420 925.941 & 1646.9   \\       
    3 & 11 &  9 & 10 &  8 & 420 929.595 & 1371.1   \\       
    2 & 11 & 11 & 10 & 10 & 420 931.991 & 1492.6   \\       
    3 & 11 & 10 & 10 &  9 & 420 938.778 & 1499.0   \\       
    2 & 11 & 10 & 10 &  9 & 420 938.985 & 1351.5   \\       
    3 & 11 & 11 & 10 & 10 & 420 945.191 & 1640.4   \\       
    2 & 11 &  9 & 10 &  8 & 420 948.726 & 1223.6   \\       
    3 & 11 & 12 & 10 & 11 & 420 950.546 & 1794.9   \\       
    3 & 11 & 13 & 10 & 12 & 420 956.037 & 1962.7   \\       
    3 & 11 & 14 & 10 & 13 & 420 962.823 & 2143.9   \\       
\end{longtable}

\begin{longtable}{cccccrr}
\caption{Calculated frequencies and intensities of $\Delta$\textit{F} = $\Delta$\textit{N} rotational transitions in the \textit{X}\textsuperscript{2}$\Sigma$\textsuperscript{+} (v=1) ground state of AlO. The constants from Goto et al. (1994) have been used. The intensities are calculated according to a Hund's case (b$\beta$$_{J}$) - (b$\beta$$_{J}$) formula (B1) from Bacis et al. (1973), at a temperature of 230 K.}\\
\hline\hline
G & N' & F' & N" & F" & Freq/MHz & Int. (230 K) \\
\hline
\endfirsthead
\caption{continued.}\\
\hline\hline
G & N' & F' & N" & F" & Freq/MHz & Int. (230 K) \\
\hline
\endhead
\hline
\endfoot
    3 &  1 &  4 &  0 &  3 &  37 916.433 &    6.0   \\       
    2 &  1 &  3 &  0 &  2 &  37 940.955 &    0.0   \\       
    2 &  2 &  2 &  1 &  1 &  75 811.789 &    0.0   \\       
    3 &  2 &  4 &  1 &  3 &  75 851.618 &   17.7   \\       
    3 &  2 &  5 &  1 &  4 &  75 865.649 &   28.9   \\       
    2 &  2 &  4 &  1 &  3 &  75 868.870 &   11.8   \\       
    2 &  2 &  3 &  1 &  2 &  75 877.360 &    4.1   \\       
    3 &  2 &  3 &  1 &  2 &  75 973.826 &    9.5   \\       
    2 &  3 &  2 &  2 &  1 & 113 785.454 &    5.5   \\       
    2 &  3 &  3 &  2 &  2 & 113 793.040 &   14.1   \\       
    2 &  3 &  1 &  2 &  0 & 113 793.103 &    0.0   \\       
    3 &  3 &  5 &  2 &  4 & 113 794.111 &   55.3   \\       
    2 &  3 &  5 &  2 &  4 & 113 799.247 &   43.0   \\       
    3 &  3 &  6 &  2 &  5 & 113 803.169 &   76.2   \\       
    2 &  3 &  4 &  2 &  3 & 113 803.475 &   26.4   \\       
    3 &  3 &  4 &  2 &  3 & 113 815.224 &   38.4   \\       
    3 &  3 &  3 &  2 &  2 & 113 841.408 &   25.1   \\       
    3 &  3 &  2 &  2 &  1 & 113 857.193 &   15.1   \\       
    2 &  4 &  3 &  3 &  2 & 151 726.258 &   33.2   \\       
    2 &  4 &  6 &  3 &  5 & 151 728.377 &  100.8   \\       
    2 &  4 &  4 &  3 &  3 & 151 728.480 &   50.8   \\       
    3 &  4 &  6 &  3 &  5 & 151 728.774 &  121.9   \\       
    2 &  4 &  2 &  3 &  1 & 151 730.961 &   19.9   \\       
    2 &  4 &  5 &  3 &  4 & 151 731.788 &   73.1   \\       
    3 &  4 &  1 &  3 &  0 & 151 735.211 &   10.3   \\       
    3 &  4 &  7 &  3 &  6 & 151 735.526 &  155.0   \\       
    3 &  4 &  5 &  3 &  4 & 151 735.995 &   94.1   \\       
    3 &  4 &  4 &  3 &  3 & 151 746.268 &   71.1   \\       
    3 &  4 &  2 &  3 &  1 & 151 750.044 &   38.8   \\       
    3 &  4 &  3 &  3 &  2 & 151 752.710 &   52.6   \\       
    2 &  5 &  7 &  4 &  6 & 189 654.765 &  192.3   \\       
    2 &  5 &  4 &  4 &  3 & 189 656.532 &   88.1   \\       
    2 &  5 &  5 &  4 &  4 & 189 656.893 &  116.7   \\       
    3 &  5 &  2 &  4 &  1 & 189 657.138 &   48.1   \\       
    2 &  5 &  6 &  4 &  5 & 189 657.809 &  151.2   \\       
    3 &  5 &  7 &  4 &  6 & 189 658.131 &  224.8   \\       
    2 &  5 &  3 &  4 &  2 & 189 660.290 &   65.3   \\       
    3 &  5 &  6 &  4 &  5 & 189 660.900 &  183.6   \\       
    3 &  5 &  8 &  4 &  7 & 189 663.606 &  272.4   \\       
    3 &  5 &  5 &  4 &  4 & 189 665.779 &  148.5   \\       
    3 &  5 &  3 &  4 &  2 & 189 666.704 &   96.1   \\       
    3 &  5 &  4 &  4 &  3 & 189 668.783 &  119.3   \\       
    2 &  6 &  8 &  5 &  7 & 227 577.375 &  324.3   \\       
    3 &  6 &  3 &  5 &  2 & 227 577.836 &  114.5   \\       
    2 &  6 &  7 &  5 &  6 & 227 580.197 &  267.6   \\       
    2 &  6 &  6 &  5 &  5 & 227 580.377 &  218.5   \\       
    2 &  6 &  5 &  5 &  4 & 227 580.899 &  176.8   \\       
    3 &  6 &  8 &  5 &  7 & 227 582.496 &  370.5   \\       
    3 &  6 &  7 &  5 &  6 & 227 583.325 &  313.6   \\       
    2 &  6 &  4 &  5 &  3 & 227 584.223 &  142.1   \\       
    3 &  6 &  4 &  5 &  3 & 227 585.035 &  187.0   \\       
    3 &  6 &  6 &  5 &  5 & 227 585.713 &  264.2   \\       
    3 &  6 &  5 &  5 &  4 & 227 587.022 &  221.9   \\       
    3 &  6 &  9 &  5 &  8 & 227 587.174 &  435.0   \\       
    3 &  7 &  4 &  6 &  3 & 265 494.808 &  216.4   \\       
    2 &  7 &  9 &  6 &  8 & 265 495.305 &  503.5   \\       
    2 &  7 &  8 &  6 &  7 & 265 497.970 &  428.9   \\       
    2 &  7 &  7 &  6 &  6 & 265 498.736 &  363.1   \\       
    2 &  7 &  6 &  6 &  5 & 265 499.743 &  305.8   \\       
    3 &  7 &  5 &  6 &  4 & 265 500.733 &  318.0   \\       
    3 &  7 &  8 &  6 &  7 & 265 501.362 &  490.9   \\       
    3 &  7 &  9 &  6 &  8 & 265 501.526 &  565.7   \\       
    3 &  7 &  7 &  6 &  6 & 265 502.401 &  424.8   \\       
    3 &  7 &  6 &  6 &  5 & 265 502.741 &  367.1   \\       
    2 &  7 &  5 &  6 &  4 & 265 502.824 &  256.9   \\       
    3 &  7 & 10 &  6 &  9 & 265 505.661 &  649.3   \\       
    3 &  8 &  5 &  7 &  4 & 303 406.685 &  360.6   \\       
    2 &  8 & 10 &  7 &  9 & 303 407.706 &  736.2   \\       
    2 &  8 &  9 &  7 &  8 & 303 410.250 &  641.4   \\       
    2 &  8 &  8 &  7 &  7 & 303 411.355 &  556.7   \\       
    3 &  8 &  6 &  7 &  5 & 303 411.841 &  495.8   \\       
    2 &  8 &  7 &  7 &  6 & 303 412.650 &  481.7   \\       
    3 &  8 &  7 &  7 &  6 & 303 413.888 &  561.4   \\       
    3 &  8 &  9 &  7 &  8 & 303 413.915 &  721.8   \\       
    3 &  8 &  8 &  7 &  7 & 303 414.147 &  636.7   \\       
    3 &  8 & 10 &  7 &  9 & 303 414.638 &  816.7   \\       
    2 &  8 &  6 &  7 &  5 & 303 415.571 &  416.3   \\       
    3 &  8 & 11 &  7 & 10 & 303 418.377 &  921.5   \\       
    3 &  9 &  6 &  8 &  5 & 341 312.432 &  553.4   \\       
    2 &  9 & 11 &  8 & 10 & 341 313.756 & 1028.6   \\       
    2 &  9 & 10 &  8 &  9 & 341 316.193 &  911.6   \\       
    3 &  9 &  7 &  8 &  6 & 341 317.083 &  726.7   \\       
    2 &  9 &  9 &  8 &  8 & 341 317.503 &  805.7   \\       
    2 &  9 &  8 &  8 &  7 & 341 318.976 &  710.7   \\       
    3 &  9 &  8 &  8 &  7 & 341 319.162 &  811.0   \\       
    3 &  9 &  9 &  8 &  8 & 341 319.813 &  906.2   \\       
    3 &  9 & 10 &  8 &  9 & 341 320.090 & 1012.4   \\       
    3 &  9 & 11 &  8 & 10 & 341 321.146 & 1129.5   \\       
    2 &  9 &  7 &  8 &  6 & 341 321.779 &  626.6   \\       
    3 &  9 & 12 &  8 & 11 & 341 324.576 & 1257.8   \\       
    3 & 10 &  7 &  9 &  6 & 379 211.141 &  801.2   \\       
    2 & 10 & 12 &  9 & 11 & 379 212.637 & 1386.7   \\       
    2 & 10 & 11 &  9 & 10 & 379 214.978 & 1245.2   \\       
    3 & 10 &  8 &  9 &  7 & 379 215.433 & 1016.9   \\       
    2 & 10 & 10 &  9 &  9 & 379 216.413 & 1116.1   \\       
    3 & 10 &  9 &  9 &  8 & 379 217.532 & 1122.0   \\       
    2 & 10 &  9 &  9 &  8 & 379 217.995 &  999.0   \\       
    3 & 10 & 10 &  9 &  9 & 379 218.444 & 1239.3   \\       
    3 & 10 & 11 &  9 & 10 & 379 219.058 & 1368.7   \\       
    3 & 10 & 12 &  9 & 11 & 379 220.313 & 1510.2   \\       
    2 & 10 &  8 &  9 &  7 & 379 220.702 &  894.0   \\       
    3 & 10 & 13 &  9 & 12 & 379 223.494 & 1664.0   \\       
    3 & 11 &  8 & 10 &  7 & 417 101.951 & 1109.8   \\       
    2 & 11 & 13 & 10 & 12 & 417 103.541 & 1816.1   \\       
    2 & 11 & 12 & 10 & 11 & 417 105.791 & 1648.2   \\       
    3 & 11 &  9 & 10 &  8 & 417 105.972 & 1372.1   \\       
    2 & 11 & 11 & 10 & 10 & 417 107.299 & 1493.7   \\       
    3 & 11 & 10 & 10 &  9 & 417 108.077 & 1500.2   \\       
    2 & 11 & 10 & 10 &  9 & 417 108.943 & 1352.5   \\       
    3 & 11 & 11 & 10 & 10 & 417 109.166 & 1641.7   \\       
    3 & 11 & 12 & 10 & 11 & 417 110.003 & 1796.3   \\       
    3 & 11 & 13 & 10 & 12 & 417 111.379 & 1964.3   \\       
    2 & 11 &  9 & 10 &  8 & 417 111.564 & 1224.5   \\       
    3 & 11 & 14 & 10 & 13 & 417 114.346 & 2145.7   \\  
\end{longtable}

\begin{longtable}{cccccrr}
\caption{Calculated frequencies and intensities of $\Delta$\textit{F} = $\Delta$\textit{N} rotational transitions in the \textit{X}\textsuperscript{2}$\Sigma$\textsuperscript{+} (v=2) ground state of AlO. The constants from Goto et al. (1994) have been used. The intensities are calculated according to a Hund's case (b$\beta$$_{J}$) - (b$\beta$$_{J}$) formula (B1) from Bacis et al. (1973), at a temperature of 230 K.}\\
\hline\hline
G & N' & F' & N" & F" & Freq/MHz & Int. (230 K) \\
\hline
\endfirsthead
\caption{continued.}\\
\hline\hline
G & N' & F' & N" & F" & Freq/MHz & Int. (230 K) \\
\hline
\endhead
\hline
\endfoot
    3 &  1 &  4 &  0 &  3 &  37 542.790 &    6.0   \\       
    2 &  1 &  3 &  0 &  2 &  37 605.344 &    0.0   \\       
    2 &  2 &  2 &  1 &  1 &  75 112.495 &    0.0   \\       
    3 &  2 &  4 &  1 &  3 &  75 137.372 &   17.7   \\       
    3 &  2 &  5 &  1 &  4 &  75 141.892 &   28.9   \\       
    2 &  2 &  4 &  1 &  3 &  75 183.222 &   11.8   \\       
    2 &  2 &  3 &  1 &  2 &  75 183.878 &    4.1   \\       
    3 &  2 &  3 &  1 &  2 &  75 266.837 &    9.5   \\       
    2 &  3 &  2 &  2 &  1 & 112 728.294 &    5.5   \\       
    2 &  3 &  1 &  2 &  0 & 112 728.397 &    0.0   \\       
    3 &  3 &  6 &  2 &  5 & 112 729.337 &   76.2   \\       
    3 &  3 &  5 &  2 &  4 & 112 729.778 &   55.3   \\       
    2 &  3 &  3 &  2 &  2 & 112 742.608 &   14.1   \\       
    3 &  3 &  4 &  2 &  3 & 112 758.605 &   38.4   \\       
    2 &  3 &  4 &  2 &  3 & 112 759.967 &   26.4   \\       
    2 &  3 &  5 &  2 &  4 & 112 763.492 &   43.0   \\       
    3 &  3 &  3 &  2 &  2 & 112 791.209 &   25.1   \\       
    3 &  3 &  2 &  2 &  1 & 112 813.425 &   15.1   \\       
    3 &  4 &  7 &  3 &  6 & 150 311.607 &  155.1   \\       
    3 &  4 &  6 &  3 &  5 & 150 314.425 &  122.0   \\       
    2 &  4 &  2 &  3 &  1 & 150 315.892 &   19.9   \\       
    2 &  4 &  3 &  3 &  2 & 150 318.652 &   33.2   \\       
    2 &  4 &  4 &  3 &  3 & 150 327.771 &   50.8   \\       
    3 &  4 &  5 &  3 &  4 & 150 329.449 &   94.1   \\       
    2 &  4 &  5 &  3 &  4 & 150 338.108 &   73.1   \\       
    2 &  4 &  6 &  3 &  5 & 150 342.478 &  100.8   \\       
    3 &  4 &  4 &  3 &  3 & 150 346.552 &   71.1   \\       
    3 &  4 &  1 &  3 &  0 & 150 358.843 &   10.3   \\       
    3 &  4 &  3 &  3 &  2 & 150 359.766 &   52.7   \\       
    3 &  4 &  2 &  3 &  1 & 150 364.690 &   38.8   \\       
    3 &  5 &  8 &  4 &  7 & 187 889.577 &  272.5   \\       
    3 &  5 &  7 &  4 &  6 & 187 893.776 &  224.8   \\       
    2 &  5 &  3 &  4 &  2 & 187 894.977 &   65.3   \\       
    2 &  5 &  4 &  4 &  3 & 187 898.631 &   88.2   \\       
    3 &  5 &  6 &  4 &  5 & 187 904.406 &  183.6   \\       
    2 &  5 &  5 &  4 &  4 & 187 905.924 &  116.7   \\       
    2 &  5 &  6 &  4 &  5 & 187 913.907 &  151.3   \\       
    3 &  5 &  5 &  4 &  4 & 187 916.259 &  148.5   \\       
    2 &  5 &  7 &  4 &  6 & 187 918.686 &  192.4   \\       
    3 &  5 &  4 &  4 &  3 & 187 926.137 &  119.4   \\       
    3 &  5 &  2 &  4 &  1 & 187 930.744 &   48.1   \\       
    3 &  5 &  3 &  4 &  2 & 187 931.560 &   96.2   \\       
    3 &  6 &  9 &  5 &  8 & 225 462.998 &  435.2   \\       
    3 &  6 &  8 &  5 &  7 & 225 468.127 &  370.7   \\       
    2 &  6 &  4 &  5 &  3 & 225 468.664 &  142.1   \\       
    2 &  6 &  5 &  5 &  4 & 225 472.704 &  176.9   \\       
    3 &  6 &  7 &  5 &  6 & 225 476.866 &  313.8   \\       
    2 &  6 &  6 &  5 &  5 & 225 479.119 &  218.6   \\       
    2 &  6 &  7 &  5 &  6 & 225 486.028 &  267.7   \\       
    3 &  6 &  6 &  5 &  5 & 225 486.299 &  264.3   \\       
    2 &  6 &  8 &  5 &  7 & 225 491.071 &  324.5   \\       
    3 &  6 &  5 &  5 &  4 & 225 494.519 &  222.0   \\       
    3 &  6 &  4 &  5 &  3 & 225 499.971 &  187.1   \\       
    3 &  6 &  3 &  5 &  2 & 225 501.378 &  114.5   \\       
    3 &  7 & 10 &  6 &  9 & 263 031.296 &  649.6   \\       
    2 &  7 &  5 &  6 &  4 & 263 036.986 &  257.0   \\       
    3 &  7 &  9 &  6 &  8 & 263 037.118 &  566.0   \\       
    2 &  7 &  6 &  6 &  5 & 263 041.225 &  305.9   \\       
    3 &  7 &  8 &  6 &  7 & 263 044.915 &  491.2   \\       
    2 &  7 &  7 &  6 &  6 & 263 047.148 &  363.2   \\       
    3 &  7 &  7 &  6 &  6 & 263 053.038 &  425.0   \\       
    2 &  7 &  8 &  6 &  7 & 263 053.483 &  429.1   \\       
    2 &  7 &  9 &  6 &  8 & 263 058.730 &  503.8   \\       
    3 &  7 &  6 &  6 &  5 & 263 060.304 &  367.3   \\       
    3 &  7 &  5 &  6 &  4 & 263 065.680 &  318.2   \\       
    3 &  7 &  4 &  6 &  3 & 263 068.245 &  216.5   \\       
    3 &  8 & 11 &  7 & 10 & 300 593.773 &  922.1   \\       
    2 &  8 &  6 &  7 &  5 & 300 599.412 &  416.5   \\       
    3 &  8 & 10 &  7 &  9 & 300 600.162 &  817.1   \\       
    2 &  8 &  7 &  7 &  6 & 300 603.763 &  481.9   \\       
    3 &  8 &  9 &  7 &  8 & 300 607.450 &  722.2   \\       
    2 &  8 &  8 &  7 &  7 & 300 609.381 &  557.0   \\       
    3 &  8 &  8 &  7 &  7 & 300 614.795 &  637.1   \\       
    2 &  8 &  9 &  7 &  8 & 300 615.388 &  641.8   \\       
    2 &  8 & 10 &  7 &  9 & 300 620.807 &  736.7   \\       
    3 &  8 &  7 &  7 &  6 & 300 621.458 &  561.7   \\       
    3 &  8 &  6 &  7 &  5 & 300 626.743 &  496.1   \\       
    3 &  8 &  5 &  7 &  4 & 300 629.976 &  360.8   \\       
    3 &  9 & 12 &  8 & 11 & 338 149.680 & 1258.7   \\       
    2 &  9 &  7 &  8 &  6 & 338 155.243 &  627.0   \\       
    3 &  9 & 11 &  8 & 10 & 338 156.564 & 1130.3   \\       
    2 &  9 &  8 &  8 &  7 & 338 159.662 &  711.2   \\       
    3 &  9 & 10 &  8 &  9 & 338 163.572 & 1013.0   \\       
    2 &  9 &  9 &  8 &  8 & 338 165.085 &  806.2   \\       
    3 &  9 &  9 &  8 &  8 & 338 170.423 &  906.8   \\       
    2 &  9 & 10 &  8 &  9 & 338 170.892 &  912.2   \\       
    2 &  9 & 11 &  8 & 10 & 338 176.469 & 1029.3   \\       
    3 &  9 &  8 &  8 &  7 & 338 176.685 &  811.5   \\       
    3 &  9 &  7 &  8 &  6 & 338 181.886 &  727.2   \\       
    3 &  9 &  6 &  8 &  5 & 338 185.533 &  553.8   \\       
    3 & 10 & 13 &  9 & 12 & 375 698.240 & 1665.2   \\       
    2 & 10 &  8 &  9 &  7 & 375 703.724 &  894.7   \\       
    3 & 10 & 12 &  9 & 11 & 375 705.582 & 1511.3   \\       
    2 & 10 &  9 &  9 &  8 & 375 708.189 &  999.8   \\       
    3 & 10 & 11 &  9 & 10 & 375 712.444 & 1369.6   \\       
    2 & 10 & 10 &  9 &  9 & 375 713.480 & 1116.9   \\       
    3 & 10 & 10 &  9 &  9 & 375 718.972 & 1240.2   \\       
    2 & 10 & 11 &  9 & 10 & 375 719.167 & 1246.1   \\       
    2 & 10 & 12 &  9 & 11 & 375 724.894 & 1387.7   \\       
    3 & 10 &  9 &  9 &  8 & 375 724.955 & 1122.8   \\       
    3 & 10 &  8 &  9 &  7 & 375 730.083 & 1017.6   \\       
    3 & 10 &  7 &  9 &  6 & 375 733.996 &  801.8   \\       
    3 & 11 & 14 & 10 & 13 & 413 238.667 & 2147.4   \\       
    2 & 11 &  9 & 10 &  8 & 413 244.076 & 1225.5   \\       
    3 & 11 & 13 & 10 & 12 & 413 246.449 & 1965.9   \\       
    2 & 11 & 10 & 10 &  9 & 413 248.572 & 1353.6   \\       
    3 & 11 & 12 & 10 & 11 & 413 253.247 & 1797.8   \\       
    2 & 11 & 11 & 10 & 10 & 413 253.773 & 1494.9   \\       
    2 & 11 & 12 & 10 & 11 & 413 259.391 & 1649.5   \\       
    3 & 11 & 11 & 10 & 10 & 413 259.559 & 1643.0   \\       
    2 & 11 & 13 & 10 & 12 & 413 265.266 & 1817.5   \\       
    3 & 11 & 10 & 10 &  9 & 413 265.343 & 1501.4   \\       
    3 & 11 &  9 & 10 &  8 & 413 270.411 & 1373.2   \\       
    3 & 11 &  8 & 10 &  7 & 413 274.505 & 1110.7   \\       
\end{longtable}

\end{document}